\documentclass[]{aa}
\pdfoutput=1
\usepackage{txfonts}
\usepackage{graphicx}
\usepackage{amssymb}
\usepackage{natbib}

\begin{document}

\title{Two Radio Supernova Remnants Discovered in the Outer
Galaxy
}

\author{Tyler J. Foster \inst{1,2} \and Brendan Cooper \inst{5} \and Wolfgang 
Reich \inst{4} \and Roland Kothes \inst{2,3} \and Jennifer West \inst{6}}

\titlerunning{Two New Supernova Remnants}
\authorrunning{Foster et al.}

\offprints{T. Foster \email{fostert@brandonu.ca}}

\institute{Department of Physics \& Astronomy,
	Brandon University
	270-18th Street, Brandon, MB
	R7A 6A9, Canada\\
\and
	National Research Council of Canada,
	Emerging Technologies -- National Science Infrastructure,
	Dominion Radio Astrophysical Observatory,
	P.O. Box 248, Penticton BC, 
	V2A 6J9, Canada\\
\and
	Dept. of Physics \& Astronomy, 
	University of British Columbia Okanagan, 
	3333 University Way, Kelowna BC 
	V1V 1V7, Canada\\
\and
	Max-Planck-Institut f\"ur Radioastronomie,
	Auf dem H\"ugel 69, 53121 Bonn, Germany\\
\and
	Dept. of Physics \& Astronomy,
	University of Calgary, 2500 University Drive NW
	Calgary, AB
	T2N 1N4, Canada\\		
\and
	Dept. of Physics \& Astronomy 
	University of Manitoba, Winnipeg, MB
	R3T 2N2, Canada
}

\date{Submitted Sept. 10, 2012}

\abstract{New and existing large-scale radio surveys of the Milky Way at 
centimetre wavelengths can play an important role in uncovering the hundreds 
of expected but missing supernova remnants in the Galaxy's interstellar medium.
We report on the discovery of two supernova remnants (SNRs) designated 
G152.4$-$2.1 and G190.9$-$2.2, using Canadian Galactic Plane Survey data.}{The 
aims of this paper are, first, to present evidence that favours the 
classification of both sources as SNRs, and, second, to describe basic 
parameters (integrated flux density, spectrum, and polarization) as well as 
properties (morphology, line-of-sight velocity, distance and physical size) to 
facilitate and motivate future observations.}{Spectral and polarization 
parameters are derived from multiwavelength data from existing radio surveys 
carried out at wavelengths between 6 and 92~cm. In particular for the 
source G152.4$-$2.1 we also use new observations at 11~cm done with the 
Effelsberg 100~m telescope. The interstellar medium around the discovered 
sources is analyzed using 1-arcminute line data from neutral hydrogen 
(\ion{H}{i}) and 45-arcsecond $^{12}$CO(J$=$1$\rightarrow$0).}{G152.4$-$2.1 is 
a barrel shaped SNR with two opposed radio-bright polarized flanks on the North 
and South. The remnant, which is elongated along the Galactic plane is evolving 
in a more-or-less uniform medium. G190.9$-$2.2 is also a shell-type remnant 
with East and West halves elongated perpendicular to the plane, and is evolving 
within a low-density region bounded by dense neutral hydrogen in the North and 
South, and molecular ($^{12}$CO) clouds in the East and West. The integrated 
radio continuum spectral indices are $-$0.65$\pm$0.05 and $-$0.66$\pm$0.05 for 
G152.4$-$2.1 and G190.9$-$2.2 respectively. Both SNRs are approximately 1~kpc 
distant, with G152.4$-$2.1 being larger (32$\times$30~pc in diameter) 
than G190.9$-$2.2 (18$\times$16~pc). These two remnants are the lowest surface 
brightness SNRs yet catalogued at 
$\Sigma_{\textrm{1GHz}}\lesssim5\times10^{-23}$~W~m$^{-2}$~Hz$^{-1}$~sr$^{-1}$.
}
{}\keywords{supernova remnants, interstellar medium}

\maketitle

\section{Introduction}
Supernovae are the predominant source of new energy and elememtal input 
into the interstellar medium (ISM). Shock waves from supernova blasts carve 
out hot ionized tunnels in the ISM, sweep up and ionize the gas in the neutral 
ISM, twist and contort the Galactic magnetic field, compress nearby 
interstellar clouds and could also trigger star formation and create
the soup of cosmic rays in the Galactic environment that plays an important 
role in pressure balance in the Milky Way ISM and star formation (among other 
things). Besides the detailed understanding of individual physical properties 
of supernova remnants (SNRs) that is needed to get a more global picture of 
their place in the ISM, a simple accurate count of the number of Galactic SNRs 
is required. Assuming a mean age of radio shell-type SNRs of 
$\gtrsim$60,000~yrs \citep{frai94} and a rate of one SNe per 30-50 years in 
spiral galaxies like the Milky Way \citep[][]{koo06}, the ISM should play host 
to between one and two-thousand radio supernova remnants at any given 
epoch. However in 2009 only 274 SNRs were catalogued in the Milky Way 
\citep[][]{gree09}, and although this has increased to $\geq$310 today in light 
of recent observations \citep[see 
catalogue\footnote{http://www.physics.umanitoba.ca/snr/SNRcat/} of][]{ferr12} 
there are still many {}``missing'' SNRs, undoubtedly due to difficulties in 
identifying very faint objects. The {}``missing SNRs problem'' is also apparent 
in models of the angular distribution of \ion{H}{ii} regions and SNRs in the 
Galaxy which predict about 1000 Galactic SNRs \citep{li91}. 

Many radio-emitting SNRs descend from type Ib, Ic and II (core collapse) 
supernova events, and are more likely to be evolving in a low density 
environment created by the stellar wind(s) of their massive and powerful 
progenitor (and cluster companions). The majority of Galactic SNRs in the 
catalogue of \citet{koth06} are found in the adiabatic expansion phase, where 
energy from their shocks is being efficiently converted into radio, X-ray 
and/or optical radiation. This indicates an observational bias favouring SNRs 
that are significantly interacting with their local ISM environment, and 
suggests that many SNRs, whose shocks are still more-or-less freely expanding 
inside pre-made cavities and have not encountered the inner walls of the 
bubbles created by stellar winds, remain unobserved. Discoveries of new SNRs 
within high-sensitivity, high-resolution and wide-field radio continuum and 
polarimetric surveys of the Galactic plane \citep[for example][]{brog06,gao11} 
are becoming common now that such surveys are more widely available. This 
success at alleviating the missing SNRs problem is being used to motivate a new 
generation of surveys (e.g. GALFACTS, POSSUM, LOFAR, and surveys to be 
performed by the SKA). Their sensitivity to all angular structures and their 
wide fields allows detection and study of large, low-surface brightness objects 
in the context of their local Galactic background, while their high-resolution 
lowers the confusion limit, allowing one to distinguish small angular 
structures from large ones (and to separate them). The Canadian Galactic Plane 
Survey \citep[CGPS,][]{tayl03} is the original such survey, and probes the 
ionized, neutral, and magneto-ionic ISM over unprecedented spatial dynamic 
range. Since 2001 this dataset has enabled the discovery of 7 new SNRs in the 
1$^{\textrm{\footnotesize{st}}}$ and 2$^{\textrm{\footnotesize{nd}}}$ quadrants 
of longitude \citep[][]{koth01,koth03,koth05,tian07,kert07}, and allowed new 
insights into the SNR population in our Galaxy. In this paper we add two of the 
faintest-known SNRs to the list of catalogued remnants in our Galaxy.

The new SNR candidates in this paper were discovered in two individual CGPS 
21~cm mosaics that had been source-subtracted and subsequently smoothed to gain 
S/N and better reveal large unseen extended emission in the survey. The two low 
surface-brightness shell-type SNRs are designated G152.4$-$2.1 and G190.9$-$2.2 
(names from Galactic $\ell,b$ coordinates of their geometric centres). Flux 
densities at five radio wavelengths ($\lambda$6~cm, 11, 21, 74 and 92~cm; in
frequency $\nu=$4.8~GHz, 2.7~GHz, 1420~MHz, 408~MHz and 327~MHz respectively) 
are measured with Stokes I data. Polarized intensity maps with B-field vectors 
from Stokes Q and U maps are presented at $\lambda$6, 11, and 21~cm (for 
G190.9$-$2.2 only 6~cm and 21~cm polarization data are available). Steep 
integrated radio spectra for each of the respective 1$\fdg$5 and 1$\degr$ 
diameter shells of G152.4$-$2.1 and G190.9$-$2.2 show the predominantly 
non-thermal nature of the sources. Finally we present 1-arcminute resolution 
\ion{H}{i} and $^{12}$CO(J=1$\rightarrow$0) line channel maps towards each 
remnant and estimate LSR velocities, kinematic distances and physical diameters 
for each. 

The primary purpose of this paper is to announce the discovery of the 
objects, present evidence for their identification as SNRs and provide only the 
basic measurements that can be reliably gleaned from survey data. It is hoped 
this will aid future specialized observations of both objects, 
observations with deeper sensitivities which will be needed to embark 
on detailed studies of their physics (e.g. spatial and frequency variation of 
spectral indexes, relative amount of free-free thermal and sychrotron emission, 
age and shock velocities, etc), beyond the scope of this discovery paper.

\section{Observations}

\subsection{21~cm and 74~cm}
The 21~cm Stokes I, Q, U continuum and \ion{H}{i} line, and 74~cm continuum 
data come from the CGPS, which is described in detail by \citet{tayl03}. The 
release of the CGPS 21~cm Q and U maps are described by \citet{land10}. For our 
purpose (a qualitative look at the polarization structure of the objects) we 
use the interferometer-only Q and U data without the large-scale structures 
restored. CGPS data were observed with the 7-antenna Synthesis Telescope (ST) 
of the Dominion Radio Astrophysical Observatory (DRAO). Near G152.4$-$2.1 the 
elliptical synthesized beam of this interferometer in the continuum is 
66$\farcs$3$\times$49$\farcs$6 at 21~cm, and 3$\farcm$8$\times$2$\farcm$8 at 
74~cm. Near G190.9$-$2.2 these values become highly elliptical at 
2$\farcm$6$\times$0$\farcm$83 and 9$\farcm$1$\times$2$\farcm$9. The practice 
of observing all baselines from 617~m to 12.9~m over 144 hours per field leads 
to good sensitivity despite the relatively small 9-m diameter of the primary 
antennae: in the vicinity of G152.4$-$2.1 (G190.9$-$2.2) measured 1-sigma 
sensitivities are $\Delta$T$_{\textrm{B}}=$60~mK (40~mK) in the 21~cm 
continuum, and 0.8~K (0.4~K) at 74~cm. In 21~cm Stokes Q they are $\sim$100~mK 
(42~mK) and in Stokes U are 80~mK (41~mK). The more elongated synthesized beam 
near $\ell=$190\degr accounts for its lower noise.
Emission from structures larger than those the shortest baseline of the
ST can sample are added in from surveys by the Effelsberg 100~m, the Stockert 
25~m and the DRAO 26~m telescopes. Data from single-dishes like these are of 
particular importance to restoring the true flux density of faint extended 
objects like new SNRs, which is missed by the incomplete spatial frequency 
sampling of ST images. The CGPS data used here are two of 84 mosaics designated 
{}``MST1'' and {}``MEQ1'', which are all available to the community through the 
Canadian Astronomy Data Centre 
(CADC\footnote{http://cadc-ccda.hia-iha.nrc-cnrc.gc.ca/cgps}).

\subsection{11~cm}
11~cm data of the two new SNRs were obtained with the Effelsberg 100-m 
telescope. For G152.4$-$2.1 new observations were made between August 2006 and 
January 2007 with a two channel cooled receiver with a bandwidth of 80~MHz 
centered at 2639~MHz. The two circularly polarized 80~MHz wide channels were 
correlated by an Intermediate Frequency polarimeter, which records Stokes I, Q 
and U simultaneously. In addition, the 80~MHz bandwidth was split into 
8$\times$10~MHz wide channels to eliminate narrow-band interference and to 
measure possible large rotation measure (RM) variations. 3C286 served as the 
main calibrator with a flux density of 10.4~Jy, 9.9\% of linear polarization at
a polarization angle of 33$\degr$. We observed a 2$\fdg$4$\times$2$\fdg$4 large 
field centred on G152.4$-$2.1 by moving the telescope beam either along 
Galactic longitude or along Galactic latitude direction with a speed of 
3$\arcmin$/sec. The individual scans of each map have 2$\arcmin$ separation 
providing full sampling of the 4$\farcm$4 beam of the Effelsberg telescope at 
2639~MHz. The standard data reduction software package, based on the 
NOD2 format \citep{hasl74} was used for the Effelsberg continuum observations. 
The maps in NOD2-format were edited for interference, and {}``scanning effects'' 
were suppressed by applying the unsharp masking method \citep[][]{sofu79}. We 
ended up with nearly 15 coverages which were combined into the final map by 
using the weaving method described by \citet{emer88}. The rms-noise of 
the final maps was measured to be 2.5~mK ($\sim$1~mJy) for Stokes I and 1.5~mK 
for Stokes Q and U. Analysis of the narrow channel polarization data does not 
indicate the existence of large RM features exceeding 100~rad~m$^{-2}$ or more. 
Earlier low resolution RM data by \citet{spoe84} indicate RMs around 
0~rad/m$^{2}$ in this direction of sky. By defining the pixels at the end of 
each scan to be zero, large-scale emission exceeding the size of the map may be 
lost, which is not an issue for studying a discrete continuum source within a 
field, but might affect Stokes Q and U. 

For G190.9$-$2.2 we extracted total intensity data from the Effelsberg 11~cm 
(2695~MHz) Galactic plane survey\footnote{Available at
http://www3.mpifr-bonn.mpg.de/survey.html} described in \citet{furs90}, which 
has a sensitivity of about $\Delta$T$_{\textrm{B}}=$16~mK. For details of the 
receiver used for this survey and the antenna response of the Effelsberg 100-m 
telescope see \citet{reic84}.

\subsection{6~cm}
Data at $\lambda$6~cm ($\nu=$4.8~GHz) in Stokes I, Q and U are from the 
Sino-German polarization survey of the Galactic plane, made with the Urumqi 
25~m telescope at Nanshan Station, China \citep{sun07,gao10}. Basic parameters 
of this survey are a resolution of 9$\farcm$5, and a measured 1-sigma 
sensitivity in our data of $\Delta$T$_{\textrm{B}}\sim$1.5~mK in I and 
$\sim$0.5~mK in Q and U. We have used the original Urumqi polarization data 
without the absolute WMAP corrections. These data already include emission of 
several degrees in extent (much larger than our objects) and so are appropriate 
for polarization studies of all but the largest discrete objects.

\subsection{92~cm and Other Frequencies}\label{other}
Stokes I data for G152.4$-$2.1 were taken from the Westerbork Northern Sky 
Survey\footnote{http://www.astron.nl/wow/testcode.php} \citep[WENSS,][]{reng97} 
at 92~cm (325~MHz). As this survey only extends north of declination $+$28.5 
degrees, data for G190.9$-$2.2 is unavailable. The measured 1-sigma flux 
density variation in these data is about 3.5~mJy/beam, or 
$\Delta$T$_{\textrm{B}}\simeq$10~K with the synthesized beam of 
54$\arcsec\times$54$\arcsec /\textrm{sin}(\delta)$. Due to the restriction on 
short observing baselines for the Westerbork interferometer (about $\sim$36~m, 
or 1.5 times an antenna diameter), the WENSS survey misses flux from structures 
greater than about 1$\fdg$4 degrees in size. This will impact the 92~cm flux 
density derived for G152.4$-$2.1 only in a small way.

A faint extended source at the position of G190.9$-$2.2 was catalogued by 
\citet{kass88} at 30.9~MHz. He did not identify the object he catalogued as 
NEK~190.9$-$2.3 as a SNR. Nonetheless the orientation and size of emission 
contours in his maps are similar to G190.9$-$2.2, and it is likely these are 
one and the same object. The integrated flux density listed by Kassim for 
NEK~190.9$-$2.3 is S$_{30.9\textrm{MHz}}=$13.6~Jy$\pm$20\%.

Dust emission from the objects was searched for in IRAS 12,~25,~60 and 
100~$\mu$m, and MSX 8~$\mu$m infrared survey data, but no emission spatially 
correlated to either source's radio morphology was found at any of these 
infrared wavelengths. The combined Wisconsin H-Alpha Mapper (WHAM) and 
Virgina Tech Spectral line Survey (VTSS) map made by \citet{fink03} was also 
searched for H$\alpha$ emission from each object (angular resolution 
$\sim$6$\arcmin$). We describe the discovery of emission corresponding with
the radio appearance of G190.9$-$2.2 in Sec.\ref{g190}.

\begin{table}
\begin{center}
\caption{Integrated flux and spectral properties of G152.4$-$2.1 and 
G190.9$-$2.2 from point-source-subtracted Stokes I maps at five radio 
frequencies.}
\label{flux}
\centering
\begin{tabular}{lr@{$\pm$}lr@{$\pm$}lr@{$\pm$}lr@{$\pm$}lc}
   \hline
   Flux ($\pm\Delta$S)& 
   \multicolumn{2}{c}{G152.4$-$2.1} &
   \multicolumn{2}{c}{G190.9$-$2.2} \\
   \hline
   S$_{4812}$[Jy] & 1.19 & 0.15 & 0.49 & 0.07\\
   S$_{2639/2695}$[Jy] & 2.07 & 0.28 & 0.57 & 0.10\\
   S$_{1420}$[Jy] & 3.16 & 0.75 & 1.21 & 0.20\\
   S$_{408}$[Jy] & 5.85 & 1.39 & 1.74 & 0.51\\	
   S$_{327}$[Jy] & 7.18 & 1.61    \\
\hline
   $\alpha$ (S$\propto\nu^{\alpha}$) & $-$0.65 & 0.05 & $-$0.66 & 0.05\\
   S$_{\nu=\textrm{1GHz}}$[Jy] & 3.54 & 1.81 & 1.30 & 0.63\\
   $\Sigma_{\textrm{1GHz}}$[10$^{-23}$Wm$^{-2}$Hz$^{-1}$sr$^{-1}$] & 5.32 & 2.71 &
   4.11 & 2.00\\
\hline
   Centre ($\ell,b$)& 
   \multicolumn{2}{c}{152$\fdg$57,$-$2$\fdg$04}&
   \multicolumn{2}{c}{190$\fdg$97,$-$2$\fdg$13}\\
   Size $\theta_{\textrm{maj}} \times \theta_{\textrm{min}}\times$angle& 
   \multicolumn{2}{c}{99$\farcm 6\times 94\farcm$8$\times$0$\degr$}&
   \multicolumn{2}{c}{69$\farcm$3$\times$60$\farcm$4$\times$30$\degr$}\\ 
\end{tabular}
\end{center}
\end{table}


\section{Appearance, Integrated Flux Densities and Spectra}
To better represent extended continuum emission, unresolved 
{}``point'' sources were first removed from each map. For 21, 74 and 92~cm 
data, sources down to Signal/Noise$\sim$2$\sigma$ were readily found by 
automated source finder {}``findsrc'', which examines an image using a matched 
{}``point-source'' wavelet filter to enhance point-like sources and locate them 
in an image. Sources are modelled with two-dimensional elliptical Gaussians 
plus a base level using the program {}``fluxfit''. Both routines are part of 
the DRAO Export Software Package, a modular software package of general and 
telescope-specific programs for processing and analysis of images made by the 
DRAO Synthesis Telescope \citep[][]{higg97}. 6 and 11~cm maps had 
relatively fewer unresolved sources since their resolution is much poorer, so 
the brightest individual sources were identified using the original 1-arcminute 
21~cm map overlaid as a guide, and these were modelled and subtracted with 
fluxfit. For brevity we do not list derived source fluxes and coordinates in 
this paper. To aid in identifying the shell and its edges the 21, 74 and 92~cm 
maps were next smoothed to 4$\farcm$4 (the resolution of the 11~cm maps). Noise 
for these final smoothed maps are 25~mK, 0.4~K, and 1.6~K (at 21, 74 and 92~cm 
respectively) for G152.4$-$2.1, while the noise level for G190.9$-$2.2 
is 15~mK and 0.4~K at 21 and 74~cm respectively. It is these source-removed and
smoothed maps that are shown in Figures \ref{g152_maps} and \ref{g190_maps}.
White 21~cm contours are overlaid on each map for comparison, also derived from 
the smoothed 4$\farcm$4 resolution map.

\subsection{G152.4$-$2.1}\label{g152}

\begin{figure*}[!ht]
\vspace{-2.6cm}
\begin{center}$
\begin{array}{cc}
\includegraphics[scale=0.36]{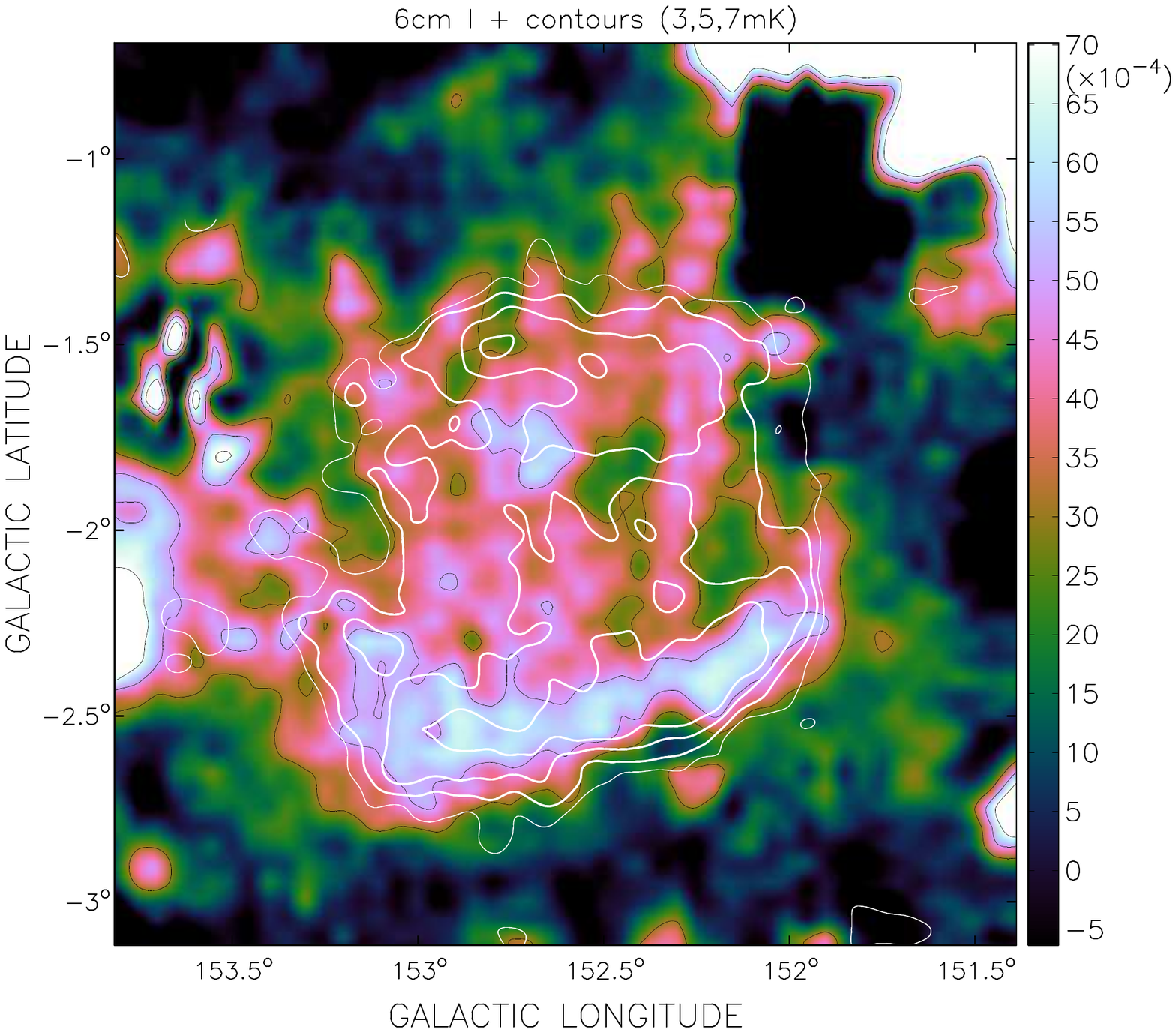}&
\hspace{-0.5cm}\includegraphics[scale=0.36]{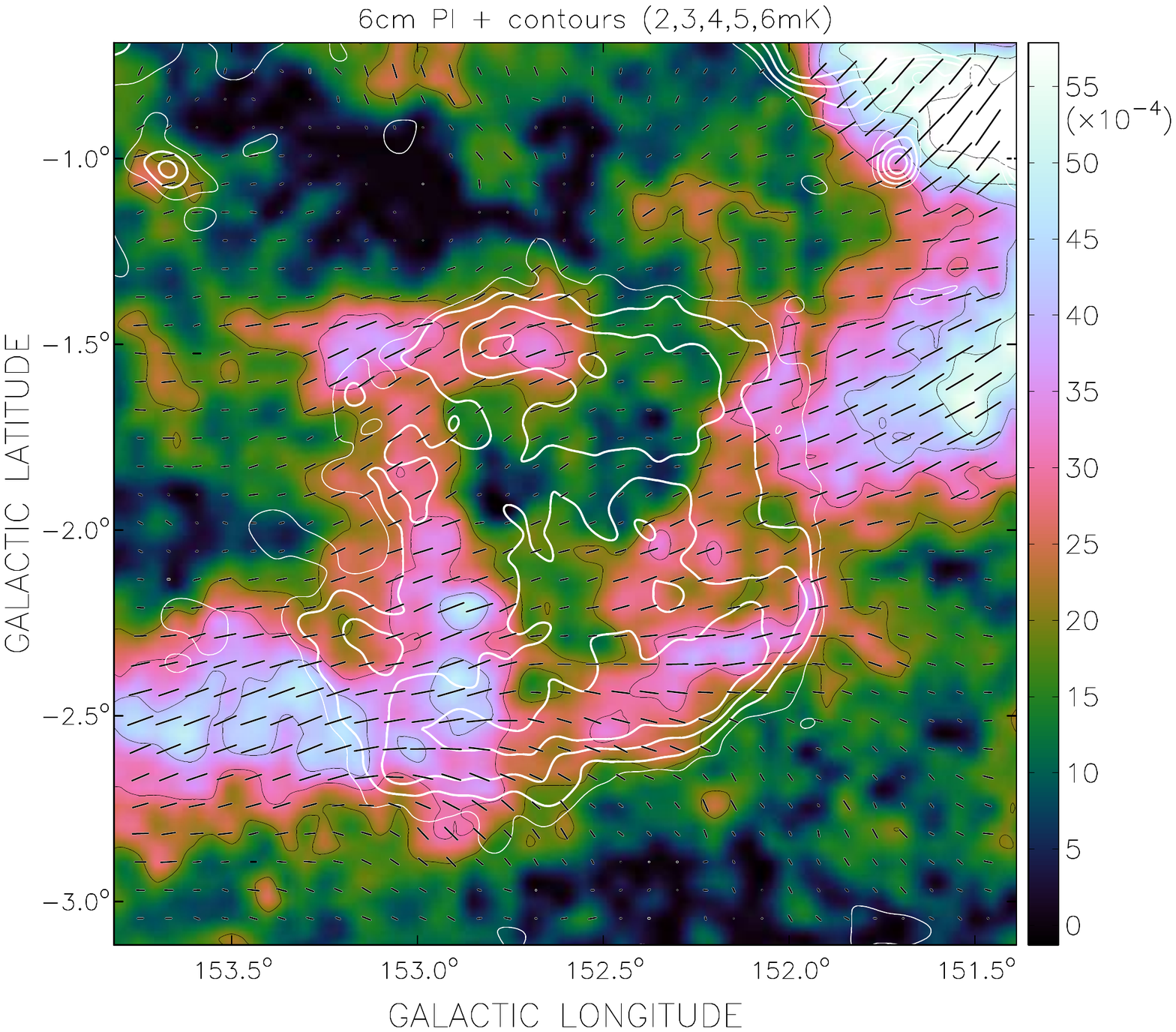}\vspace{-3.5cm}\\
\includegraphics[scale=0.36]{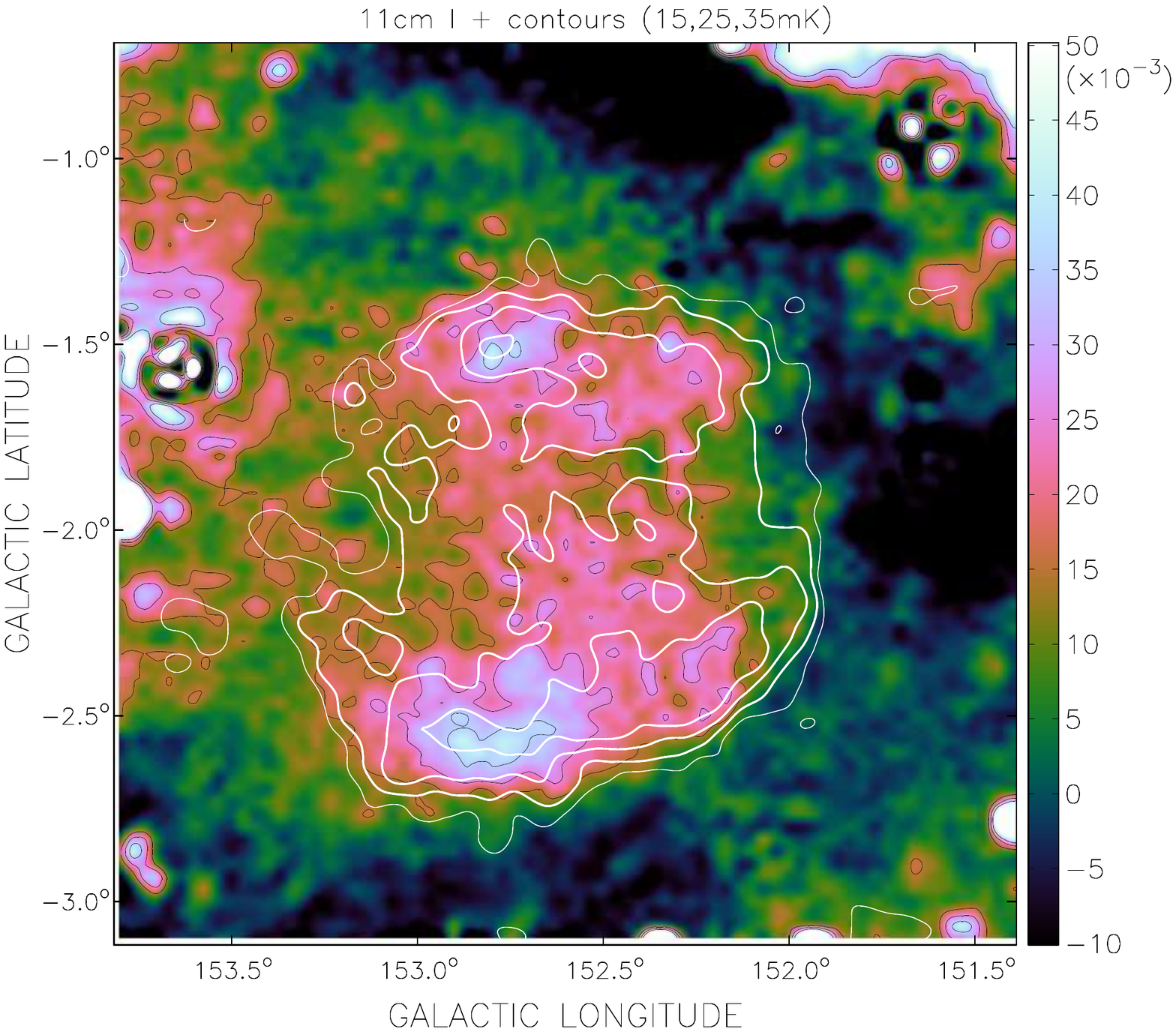}&
\hspace{-0.5cm}\includegraphics[scale=0.36]{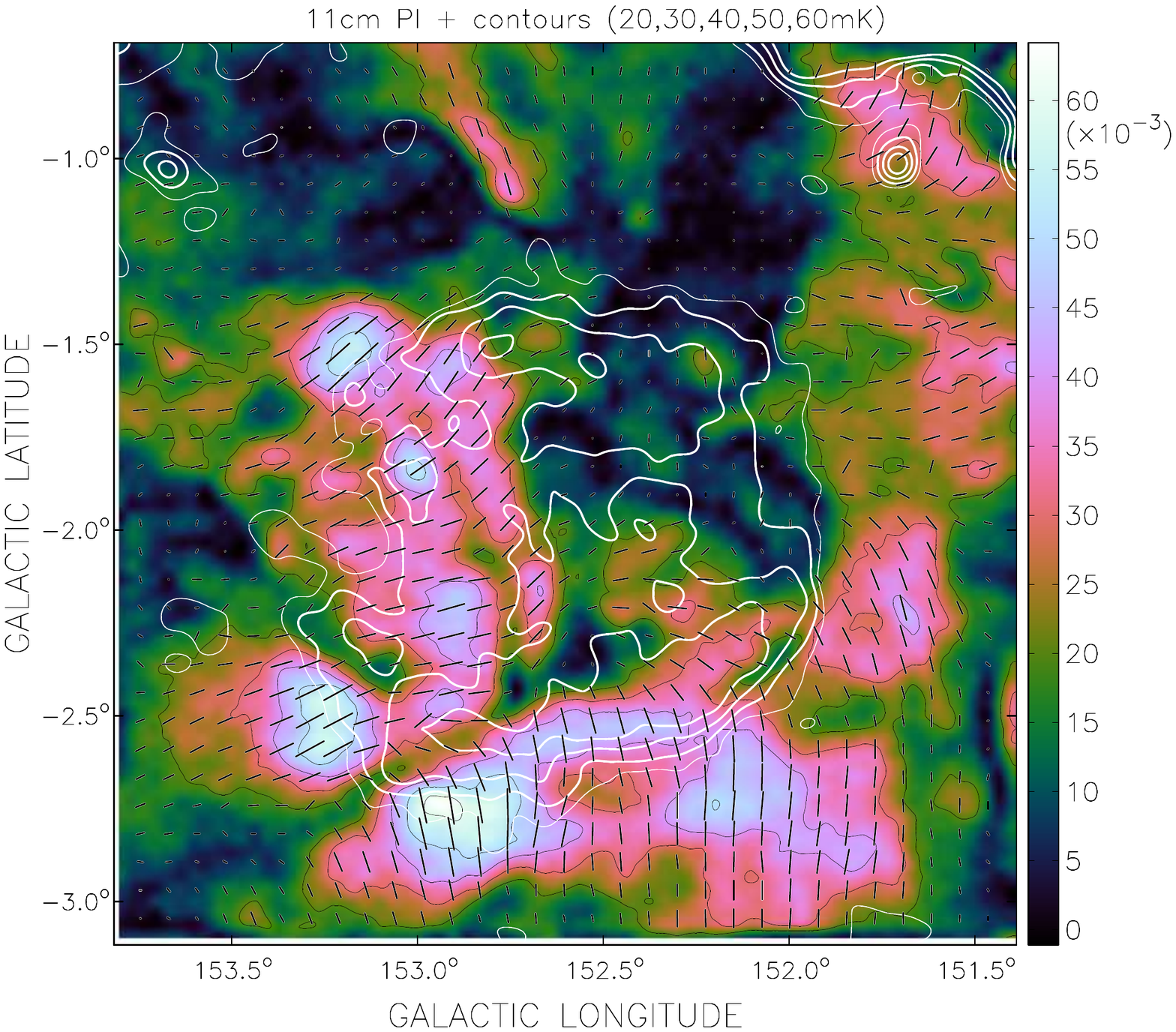}\vspace{-3.5cm}\\
\includegraphics[scale=0.36]{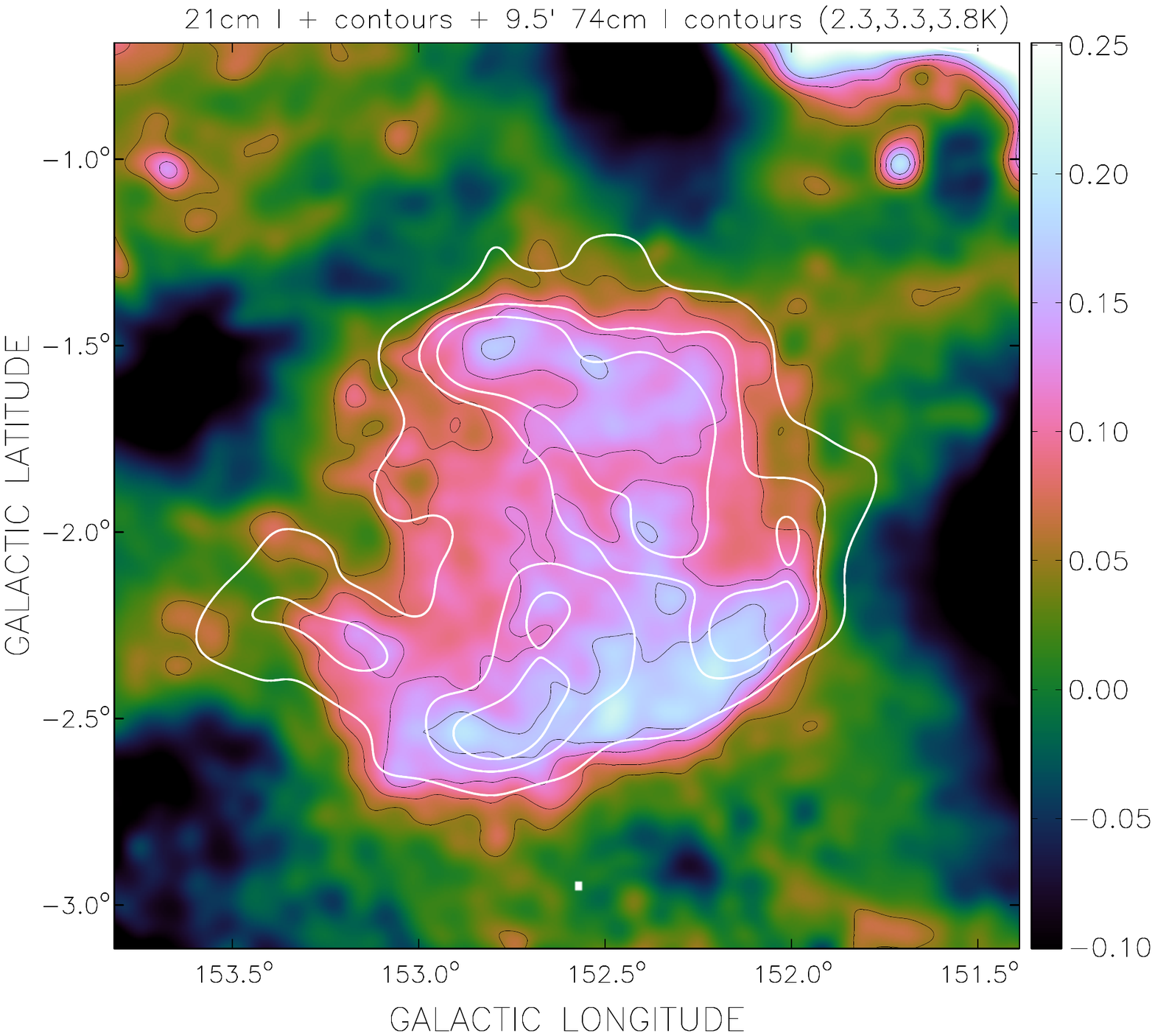}&
\hspace{-0.5cm}\includegraphics[scale=0.36]{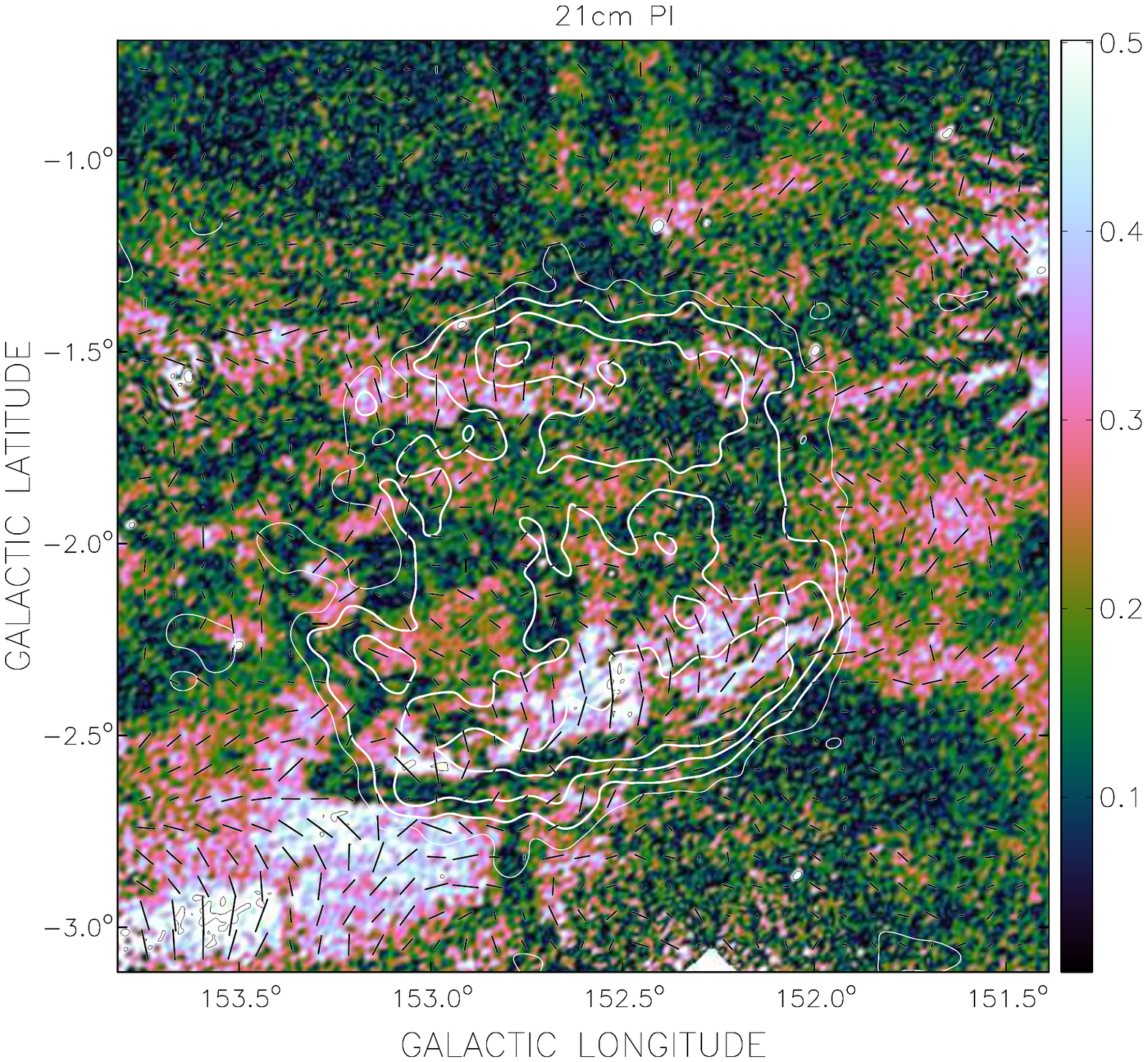}\vspace{-3.5cm}\\
\includegraphics[scale=0.36]{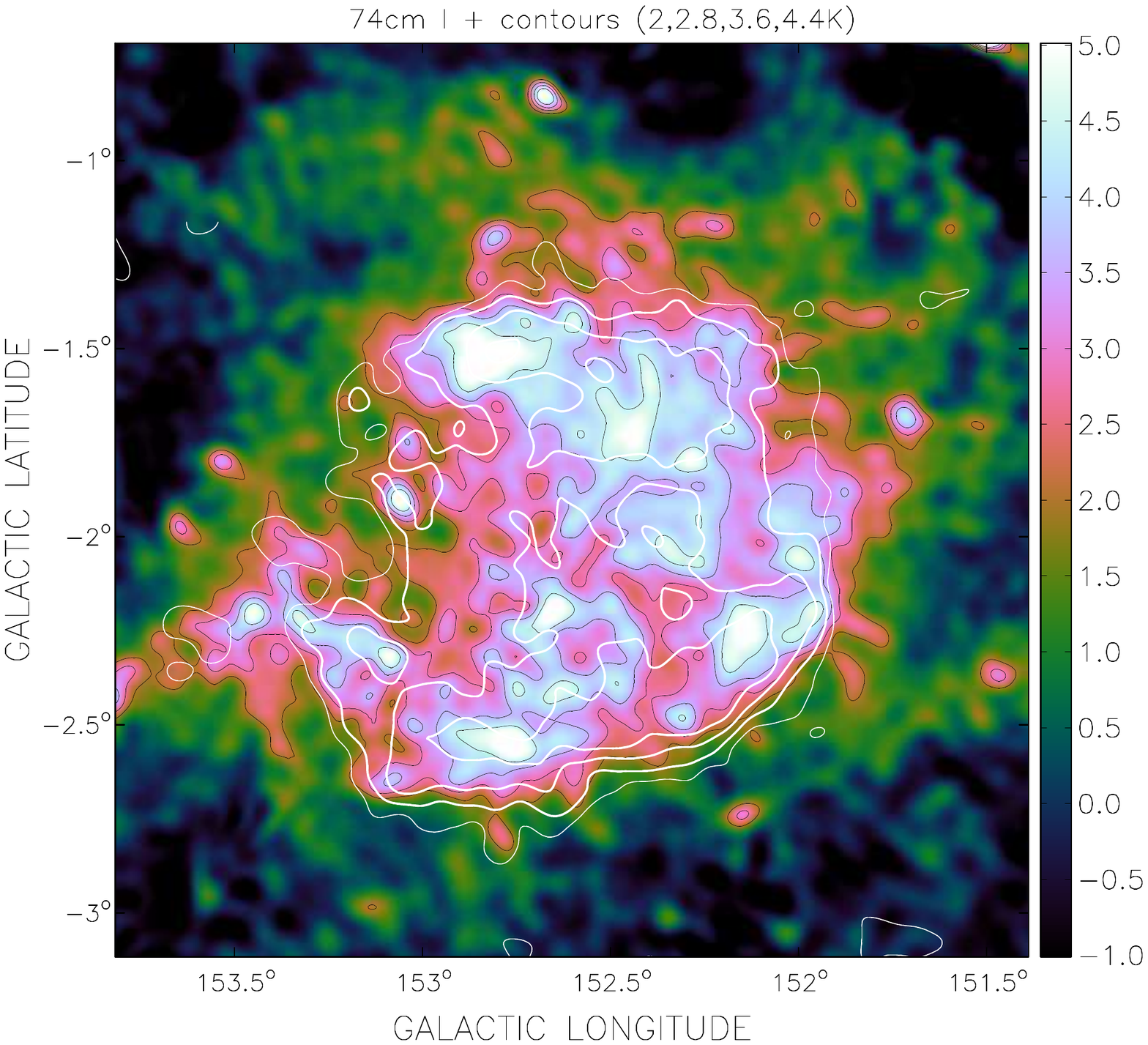}&
\hspace{-0.5cm}\includegraphics[scale=0.36]{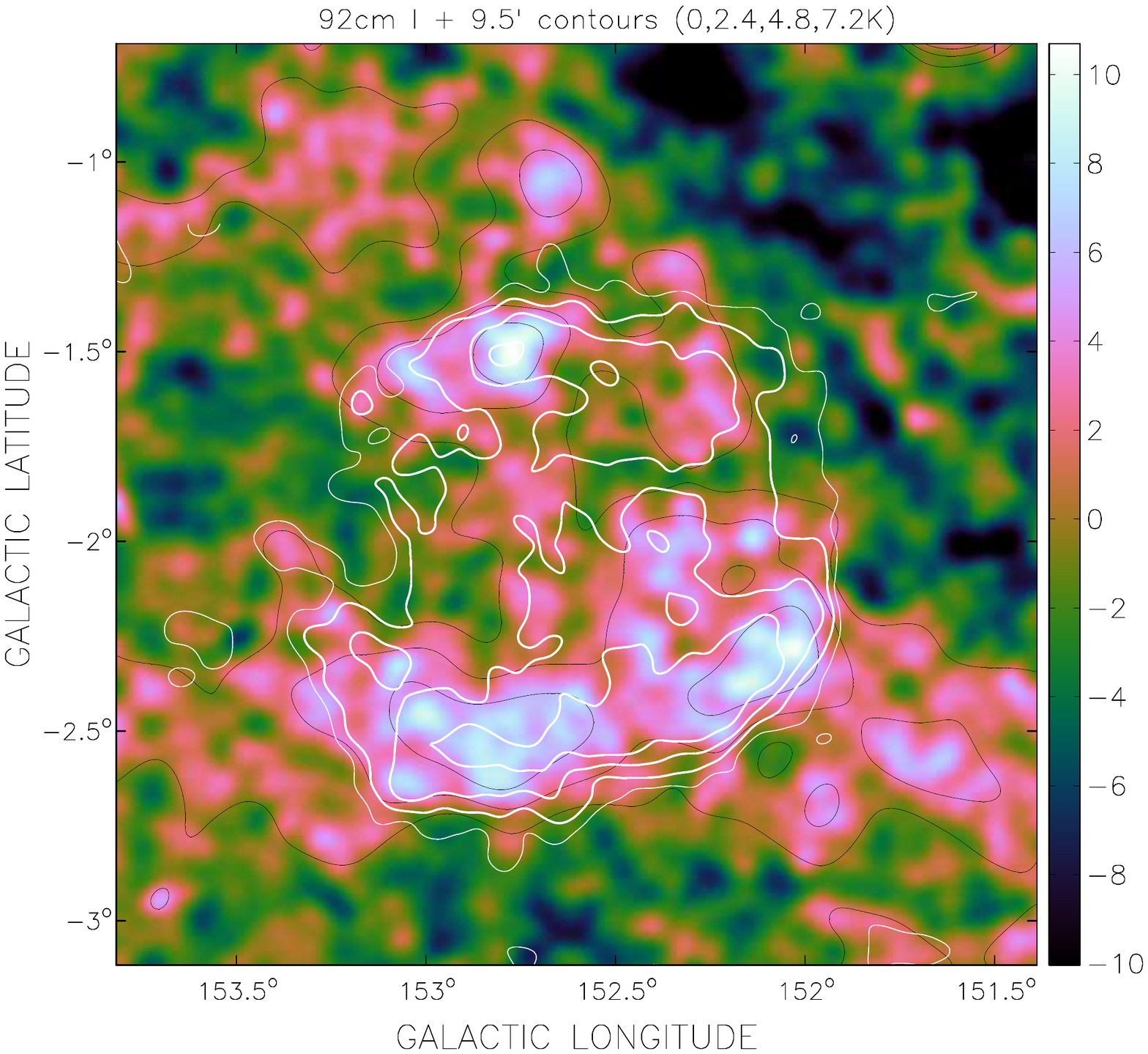}\vspace{-1.6cm}\\
\end{array}$
\caption{Total (I) and polarized intensity (PI) contoured maps of G152.4$-$2.1. 
The emission features (colour bars in brightness temperature units of K, at 
right) are themselves outlined in thin black contours (levels indicated in the 
text above each panel). B-field vectors are overlaid on PI maps (length 
proportional to PI; scale differs between maps). Resolutions are 9$\farcm$5 at 
6~cm and 4$\farcm$4 at 11, 21, 74 and 92~cm (except 21~cm PI which is 
1$\arcmin$). PI maps are missing large-scale backgrounds. Four 21~cm contours 
(white) at 50,80,120, and 160~mK are shown overlaid on all maps except 21~cm I, 
which has contours drawn from a smoothed 9$\farcm$5 74~cm I map. 
}
\label{g152_maps}
\end{center}
\end{figure*}

We refer here to total intensity maps at $\lambda$6~, 11, 21, 74, and 92
centimetres (4.8~GHz, 2.7~GHz, 1420~MHz, 408~MHz and 327~MHz) in Figure 
\ref{g152_maps}. All total (I) and polarized intensity (PI) maps are presented 
in the colour-scheme of \citet{gree11}. The emission structure of G152.4$-$2.1 
is best seen in the 21~cm map. The very bright \ion{H}{ii} region Sh2-206 is 
seen in the top-right corner of I maps (and 6~cm PI map). The radio emission 
from this SNR candidate consists of a pair of enhanced shells of 
emission with steep outer edges in the North and South. This is the typical 
{}``double loop'' or {}``barrel''-shaped structure of a pure shell-type 
supernova remnant \citep[see][for example]{manc87,gaen98}, where two shells of 
enhanced radio emission are found atop a more diffuse central emission plateau 
which is likely generated by those parts of the source that are expanding 
towards us or moving away from us. \citet{gaen98} found that the orientation of 
these barrel-shaped SNRs is mostly parallel to the Galactic plane - as is the 
case for G152.4$-$2.1. He suggests that the reason for this shape is a 
combination of the uniform magnetic field structure which is preferentially 
parallel to the Galactic plane and the expansion into ISM structures which are 
elongated in longitudinal direction. \citet{casw77} also describes asymmetries 
in shell remnants as the outcome of ISM density gradients.

We determine the polarization properties of G152.4$-$2.1 using Stokes Q and U 
polarization maps at 6, 11, and 21~cm. In Fig.~\ref{g152_maps} we display
polarized intensity images with B-vectors overlaid. The length of each 
vector is determined by the polarized intensity at that point; however, as the 
intention is simply to highlight the field's direction, the scale in cm per mK 
is arbitrary from map to map in Figures \ref{g152_maps} and \ref{g190_maps}. 
Synchrotron emission from supernova remnants should be inherently polarized up 
to 70\%; however Faraday rotation by the ISM and internal effects will 
depolarize this signal, and this depolarization is much more significant at 
21~cm than at 6~cm, effectively scrambling low-frequency polarized intensity 
from a SNR. This effect can be easily seen in our polarized emission images. 
While there is a strong correlation between total power and polarization at 
6~cm, the ambient polarization features become more and more dominant at longer 
wavelength.

G152.4$-$2.1 shows polarized intensity at 6~cm well correlated to continuum 
features, a good indicator that the emission is synchrotron in nature. The 
fractional polarization on the brightened regions of the shell is between 50 
and 60\,\%. G152.4$-$2.1 also seems to sit on an elongated slightly curved 
feature in polarized intensity more than 2 degrees in length. This large 
polarization filament seems to have a tangential intrinsic magnetic field as 
indicated by the B-vectors. At this high frequency, where effects of Faraday 
rotation should be small, the B-vectors of the SNR candidate indicate a 
tangential magnetic field in the northern shell, which is expected for a mature 
shell-type SNR. The southern shell shows a slight rotation of up to 30 degrees 
clockwise from a tangential structure, which requires a rotation measure of 
about $-$130~rad\,m$^{-2}$. This would produce a rotation of about 90 degrees 
for polarization angles at 11~cm and we find the B-vectors in the 11~cm
polarization image in this area to be almost radial. We did not produce a 
rotation measure map because at the lower frequencies the background/foreground 
polarization structures become more and more dominant and it is difficult to 
distinguish between them and genuine SNR emission.

From the polarized emission structure we conclude that the SNR is likely 
interacting with the large polarization filament, the nature of which is 
unclear.
Its bright polarized emission implies an enhanced, compressed internal 
magnetic field. 
In this scenario the progenitor star of G152.4$-$2.1 would have exploded at 
the upper edge of this filament, sweeping up its magnetic field explaining the filament's
gap at the location of the SNR at 6~cm and the similarity of the B-vectors of 
both objects.

The 21~cm PI map is shown in Figure~\ref{g152_maps}. The only morphological
correlation between the PI and I maps at 21~cm seems to be the distinct dark 
channel of depolarization that snakes West to East across most of the South 
shell's face, which tightly follows the continuum around the South-East 
corner of G152.4$-$2.1. It is moderately consistent with a compressed magnetic 
field aligned in the East-West direction along the shell which cancels with the 
principally North-South field seen in the unrelated bright PI from the ISM 
immediately above it. Some of this bright PI emission on the shell's Western 
face at $\ell=$152$\fdg$2,$b=-$2$\fdg$4 could originate with the SNR's shell, 
but how much is impossible to say from these maps, as is its precise role in 
creating the depolarization {}``snake''.
\subsection{G190.9$-$2.2}\label{g190}
\begin{figure*}[!ht]
\vspace{-2cm}
\begin{center}$
\begin{array}{cc}
\includegraphics[scale=0.36]{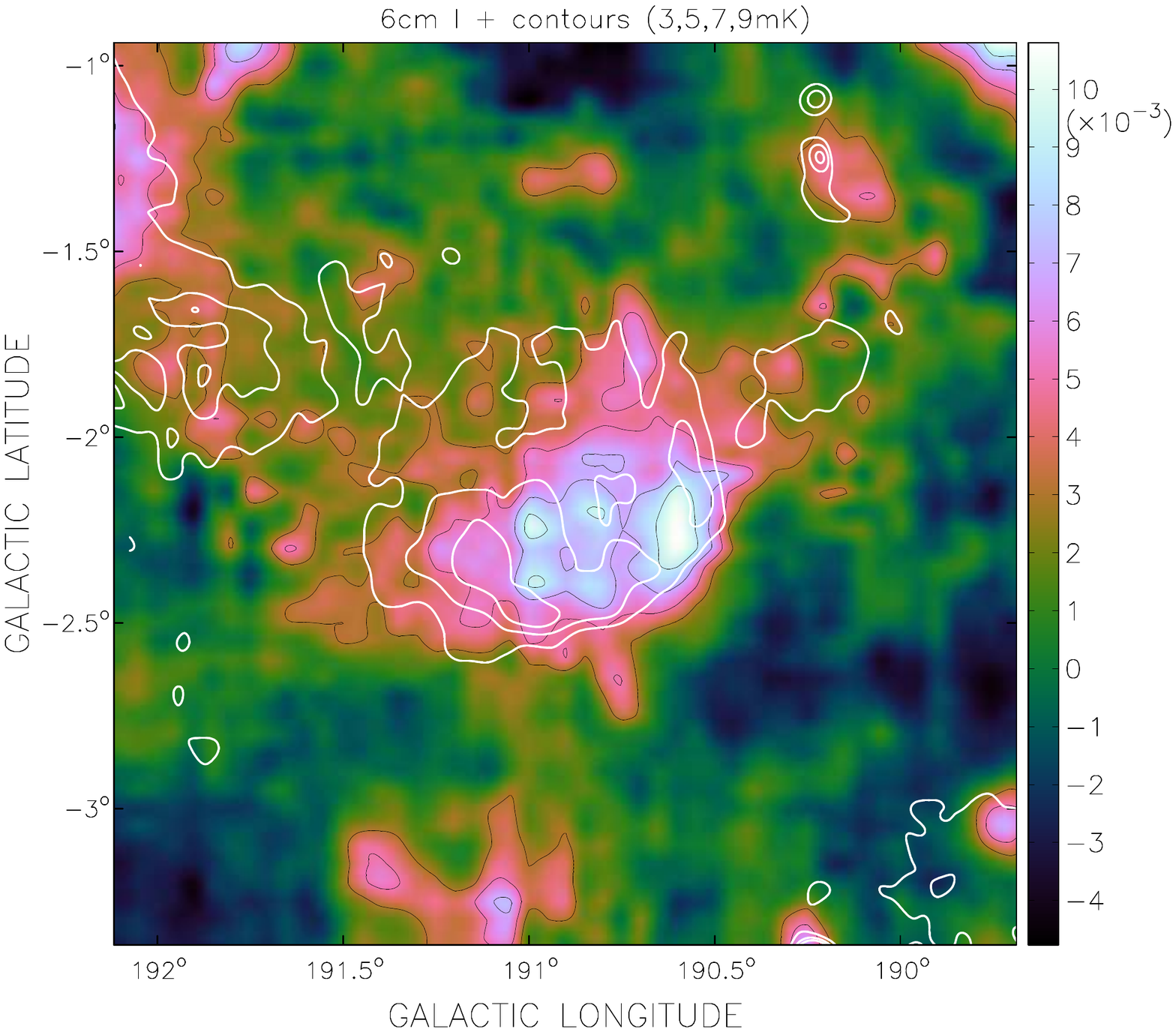}&
\hspace{-0.5cm}\includegraphics[scale=0.36]{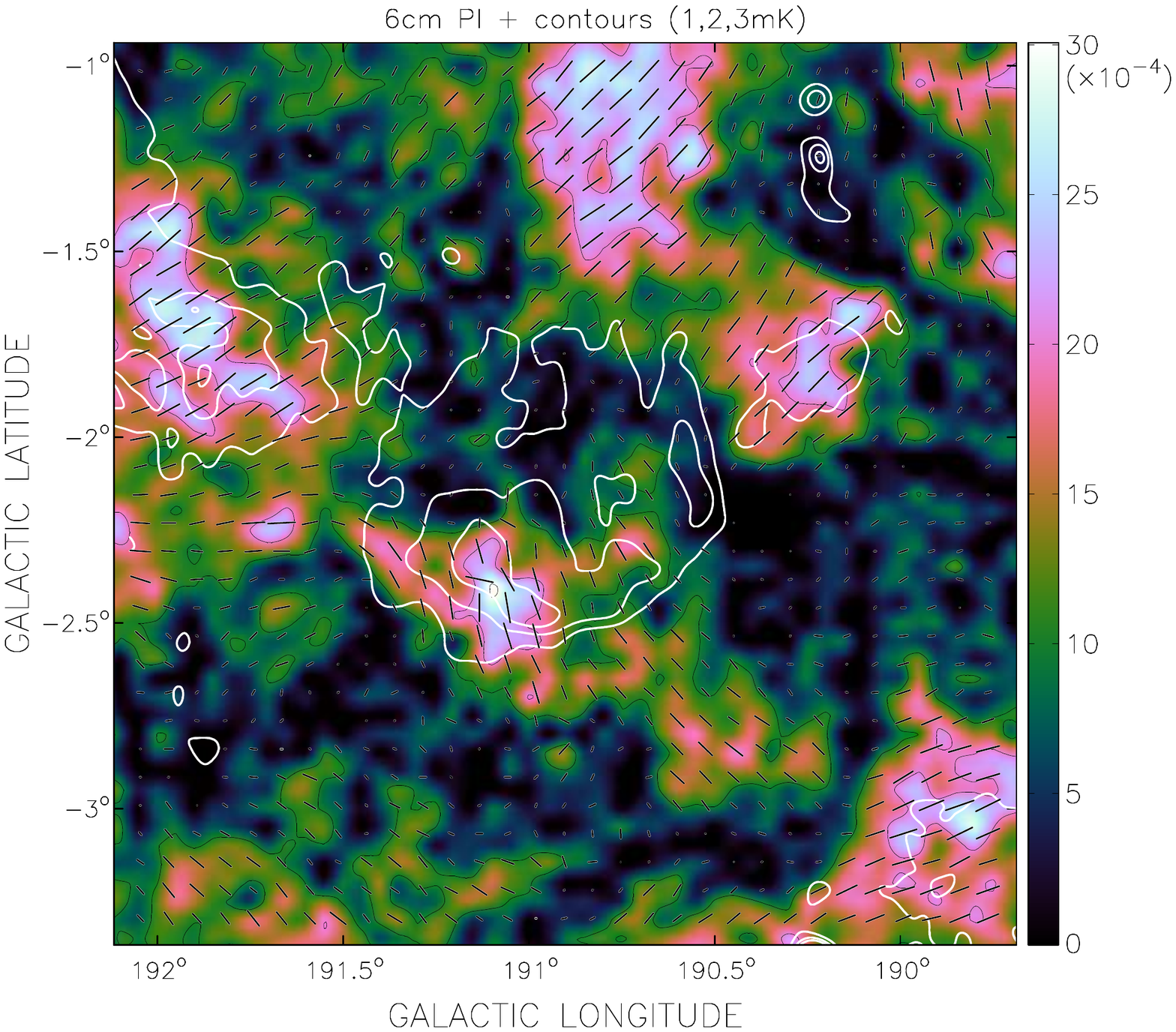}\vspace{-3.45cm}\\
\includegraphics[scale=0.36]{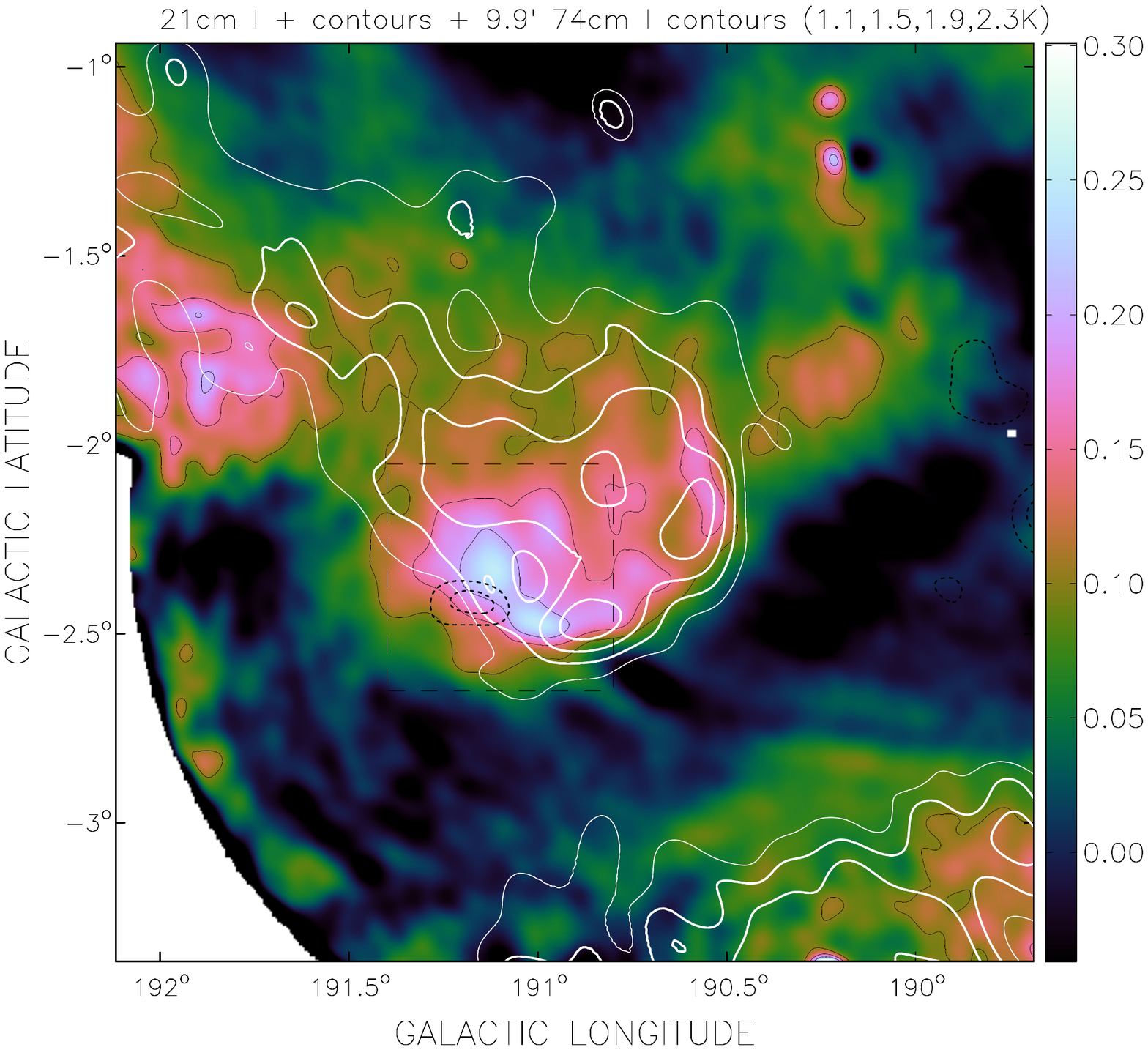}&
\hspace{-0.5cm}\includegraphics[scale=0.36]{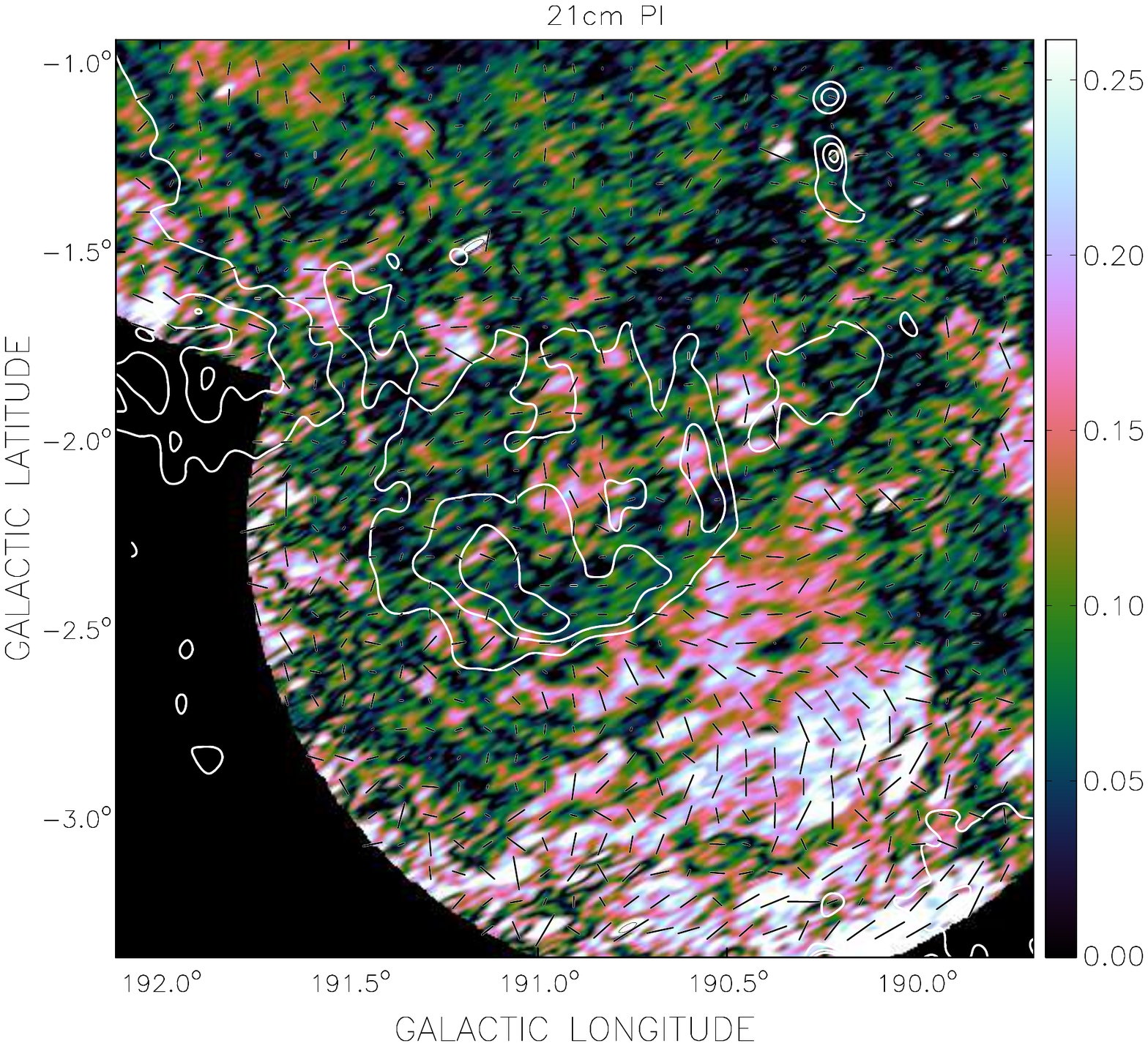}\vspace{-3.45cm}\\
\includegraphics[scale=0.36]{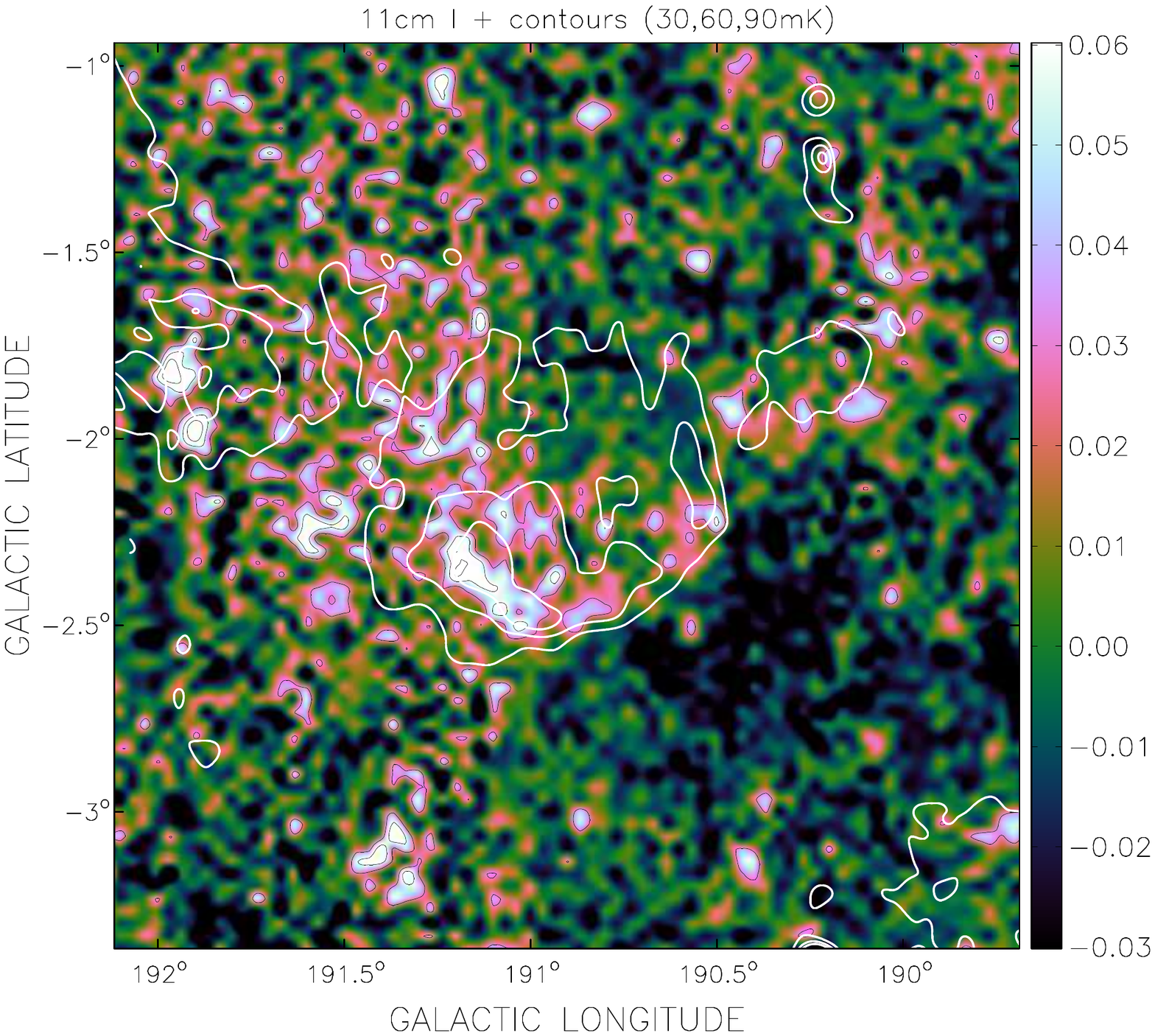}&
\hspace{-0.5cm}\includegraphics[scale=0.36]{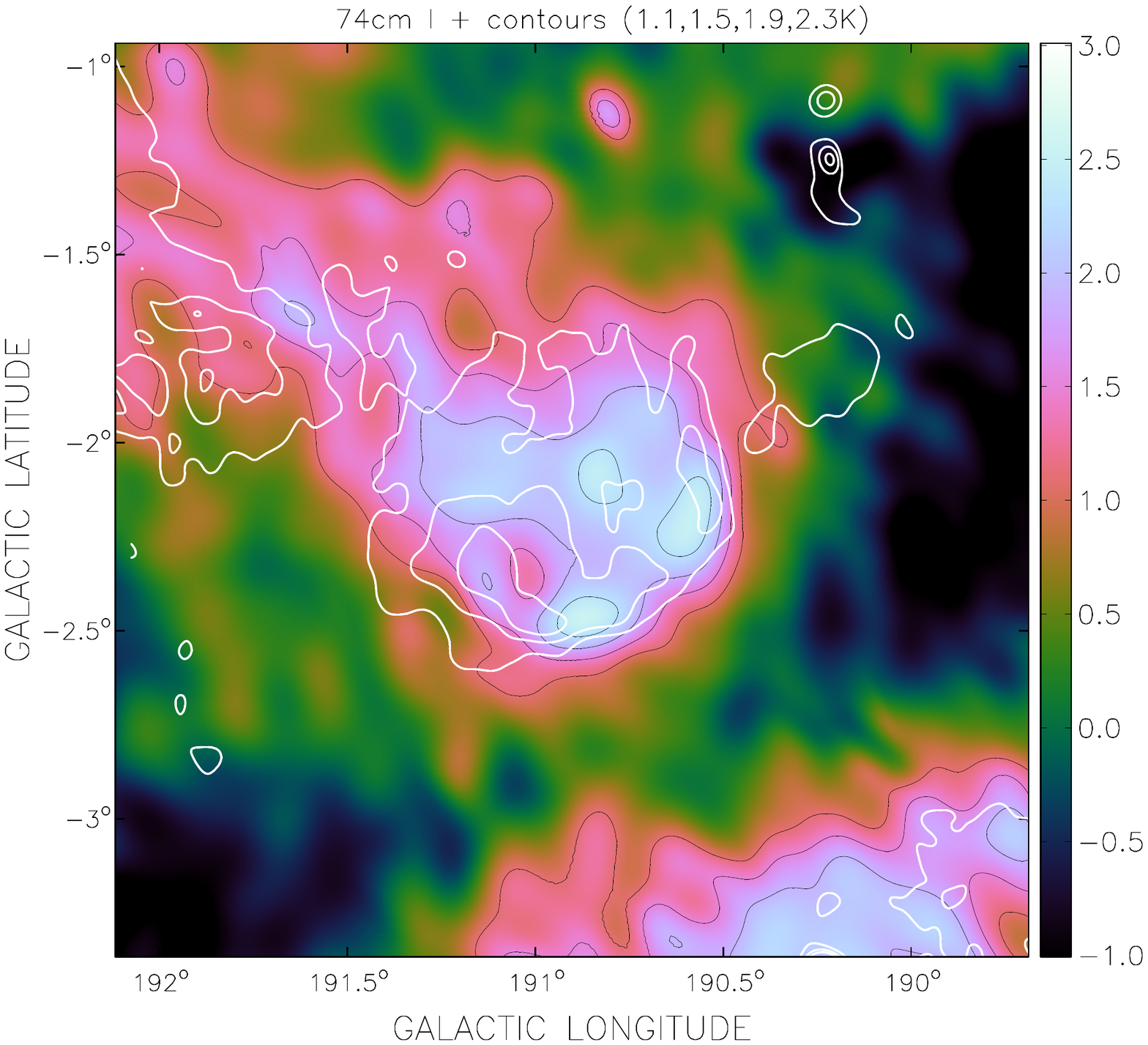}\vspace{-1.5cm}\\
\end{array}$
\caption{Total (I) and polarized intensity (PI) contoured maps of G190.9$-$2.2. 
The emission features (colour bars in brightness temperature units of K, at   
right) are themselves outlined in thin black contours (levels indicated in the 
text above each panel). B-field vectors are overlaid on PI maps (length
proportional to PI; scale differs between maps). Resolutions are 9$\farcm$5 
(6~cm), 4$\farcm$3 (11~cm, 21~cm I), 2$\farcm$6$\times$0$\farcm$85 (21~cm PI), 
and 9$\farcm$9 (74~cm). The box outline on the 21~cm panel corresponds to the 
optical image in Figure~\ref{optical}; the elliptical patch of H$\alpha$ 
emission identified in WHAM+VTSS data is also shown (dotted; 11,12~Rayleighs) 
on this panel. Three 21~cm contours (white) at 100,150, and 200~mK are shown 
overlaid on all maps, except the 21~cm panel which has white contours drawn 
from the 9$\farcm$9 74~cm I map.
}
\label{g190_maps}
\end{center}
\end{figure*}

Total Power maps at 6, 11, 21 and 74 centimetres (4.8~GHz, 2.7~GHz, 1420~MHz, 
and 408~MHz) are shown in Figure~\ref{g190_maps}. The first impression of 
G190.9$-$2.2 is one of an incomplete shell of emission with a brightened and 
sharpened Southern edge to the object. On closer inspection however we believe 
that G190.9$-$2.2 shows a barrel-shaped structure very similar to G152.4$-$2.1, 
with two lobes of enhanced shell emission on either side (East and West) of 
more diffuse emission in the centre. The Eastern lobe is seen well at 21~cm and 
11~cm, and the Western half is bright at 6~cm. The angle of the barrel's line 
of cylindrical symmetry on the sky is about 110 degrees from the Galactic 
longitude axis. If there is a correlation between this orientation of the 
``barrel'' and the ambient magnetic field and elongation of ambient structures 
as proposed by \citet{gaen98}, we must have either a slight anomaly here where 
the local magnetic field has a large angle with Galactic longitude or 
G190.9$-$2.2 is a SNR expanding inside a stellar wind bubble. Since the radio 
surface brightness of this SNR candidate is very low it is likely expanding in 
a low density environment, and in fact G190.9$-$2.2 does appear within a 
gas-poor circular region bounded by both dense molecular clouds in the East and 
West, and dense neutral hydrogen caps in the North and South (see 
Section~\ref{vlsr}).

\begin{figure}[!ht]
\vspace{-3cm}
\begin{center}$
\begin{array}{c}
\includegraphics[scale=0.4]{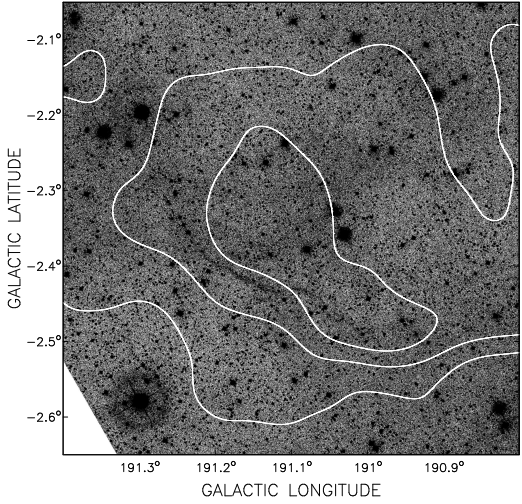}
\end{array}$
\vspace{-1.5cm}
\caption{Cutout of the bright radio shell of G190.9$-$2.2 corresponding to the 
box outlined on the 21~cm panel of Fig.~\ref{g190_maps}, showing optical (red) 
emission from the Second Palomar Observatory Sky Survey (bright emission is 
shaded dark here). A thin compressed filament of red nebulosity is seen running 
parallel and next to the innermost 21~cm radio contour, which is the completely 
enclosed white contour in the centre (the three white contours here are the 
same as are in the 21~cm panel of Figure~\ref{g190_maps}). This filament likely 
traces shocked gas downstream of the SNR shock.
}
\label{optical}
\end{center}
\end{figure}

Optical emission atop G190.9$-$2.2 is seen in WHAM+VTSS maps: a bright isolated
unresolved patch of H$\alpha$ emission some $\sim$5~Rayleighs above background 
is found to correspond with the brightest part of the 21~cm emission (the 
South-East portion of the radio shell). The patch of H$\alpha$ is shown in the 
21~cm panel of Figure~\ref{g190_maps}, outlined with dotted contours. The 
origin of this H$\alpha$ emission is clearly seen in an optical (red) POSS-II 
image in Figure \ref{optical}, the outline of which is shown as the dashed box 
atop the 21~cm panel in Fig.~\ref{g190_maps}. The patch corresponds to a very 
long, thin filament of red nebulosity that runs parallel to the 200~mK 21~cm 
contour (the innermost black contour around the brightest inner portion of the 
South-East shell on the 21~cm panel of Fig.~\ref{g190_maps}; the same contours 
are drawn in white on Fig.~\ref{optical}). This likely is highly compressed 
cooling post-shock gas just downstream of the SNR shock wave, which is 
expanding in a South-Easterly direction. 

The PI map at 6~cm (Fig.~\ref{g190_maps}) shows the eastern shell to be 30-50\% 
polarized, but the other parts of the SNR candidate seem to be unpolarized. 
This could be the result of the low surface brightness combined with uncertain 
contributions from the unrelated interstellar medium along the line-of-sight. 
The B-vectors derived for the eastern shell indicate a tangential magnetic 
field which would be expected for a mature shell-type remnant. This provides 
further evidence that G190.9$-$2.2 is a supernova remnant.

\subsection{Integrated Flux Densities}

\begin{figure}[!ht]
\begin{center}
\resizebox{\hsize}{!}{\includegraphics{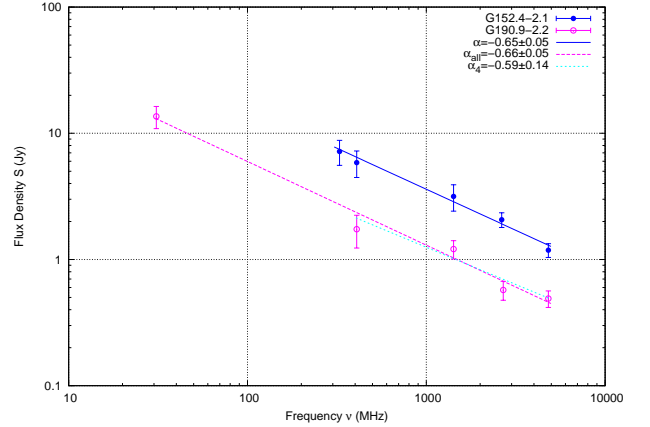}}
\caption
{Integrated radio flux spectra for each object, and overlaid power-law fitted 
with least squares weighted by uncertainties. For G190.9$-$2.2 the spectrum 
with index $\alpha_4$ is calculated over 408~MHz-4.8~GHz ($\lambda$74 - 6~cm) 
only and does not include the 30.9~MHz flux point from \citet{kass88}.}
\label{intflux}
\end{center}
\end{figure}

Integrated flux densities are summarized in Table~\ref{flux}. Tabulated fluxes 
are averages obtained over 20 polygons each with two kinds of background 
surfaces fitted to the data values at the polygon vertices: twisted-plane and 
twisted-quadratic surfaces. Twisted plane surfaces have the property that 
orthogonal cuts through them at one particular rotation angle are linear, and 
twisted quadratics have one orientation in which all cuts through the surface 
are linear, and all orthogonal cuts are quadratic. Uncertainties in fluxes 
($\Delta$S) accumulate from three sources: i) the uncertainty in the total 
intensity calibration of each map ($\pm$4,~10,~5,~15~and~20\% for 
$\lambda=$6,~11,~21,~74,~and~92~cm respectively), ii) the deviation in the 
background-subtracted flux over different polygons enclosing the emission 
(drawn by eye), and iii) the mean difference in background flux levels 
estimated with both types of background surface fits. A linear fit of 
log~S versus log$\nu$ was calculated with least squares weighted by $\Delta$S. 
Each object shows a steep power-law radio spectrum identifying the integrated 
emission as non-thermal (see Fig.~\ref{intflux}). For G152.4$-$2.1 the 
integrated spectral index is $\alpha=-$0.65$\pm$0.05 
(S$_{\nu}\propto\nu^{\alpha}$). For G190.9$-$2.2 $\alpha=-$0.66$\pm$0.05 
(without the 30.9~MHz flux point of \citet{kass88} $\alpha=-$0.59$\pm$0.14). 
Also reported in Table~\ref{flux} is each object's estimated flux density and 
surface brightness at 1~GHz, the mean centroid coordinates, and the major and 
minor angular diameters and orientation angle of each shell (fitted by eye). At 
1~GHz the surface-brightness of both SNRs (as listed in Table~\ref{flux})
would make them the faintest yet catalogued, below the previous faintest SNR 
G156.2$+$5.7 
\citep[5.8$\times$10$^{-23}$~W~m$^{-2}$~Hz$^{-1}$~sr$^{-1}$;][]{reic92}.

\section{\ion{H}{i} and CO Velocities and Distances}
\label{vlsr}
We determine systemic LSR velocities for both SNRs along their lines-of-sight 
using $\sim$1$\arcmin$ resolution CGPS 21~cm \ion{H}{i} line data and 
45$\arcsec$ resolution 2.6~mm (115.3~GHz) $^{12}$CO(J=1$\rightarrow$0) line 
data from the Exeter FCRAO CO Galactic Plane Survey \citep[described 
in][]{mott10,brun12}. 

\begin{figure*}[!ht]
\begin{center}
\includegraphics[scale=0.85]{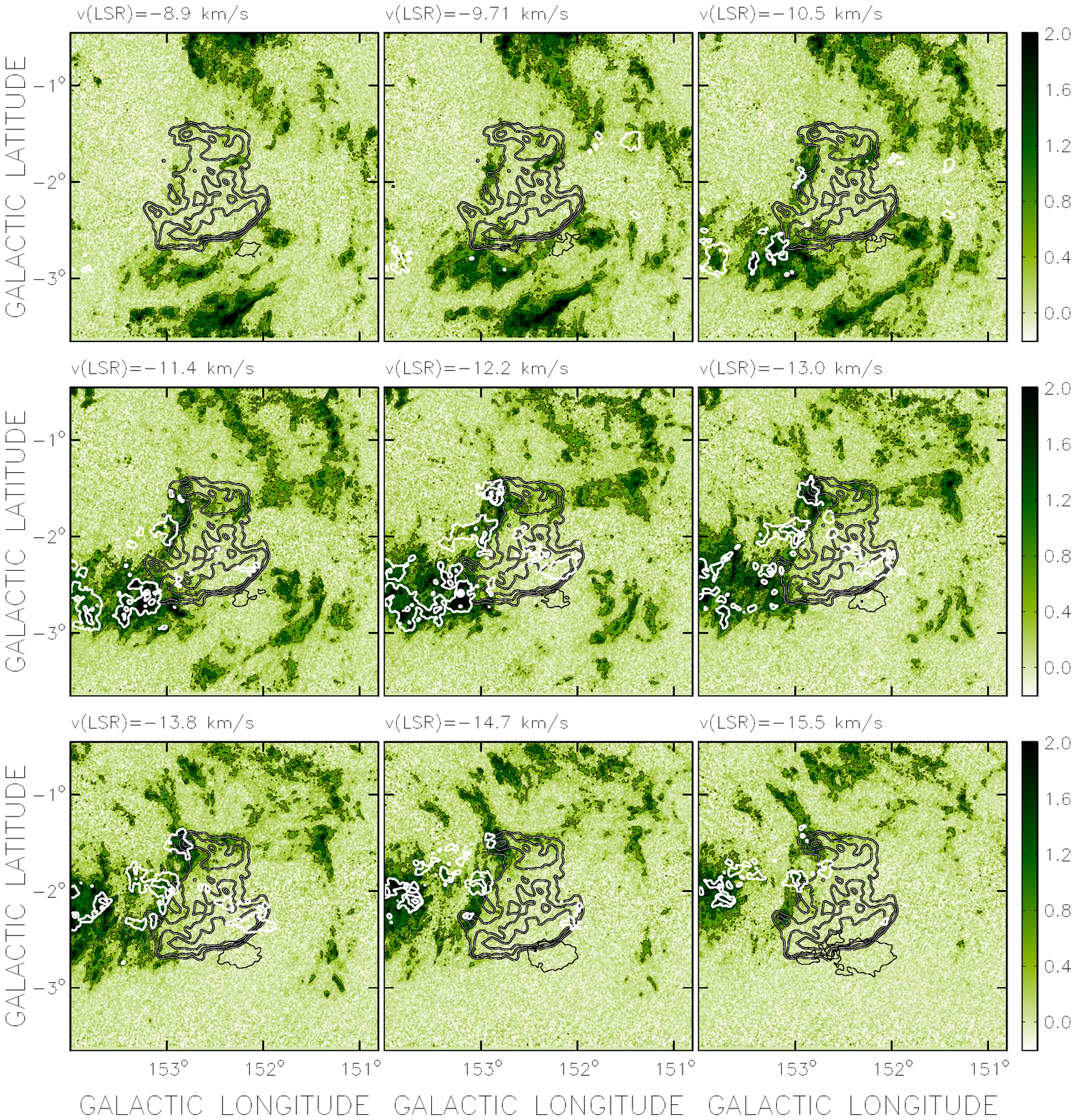}\vspace{-6cm}
\caption{$^{12}$CO line channel maps towards G152.4$-$2.1. The brightness scale 
is antenna temperature T$_{\textrm{A}}$ in K, and the angular resolution of the 
maps is 1$\arcmin$. Three thick white-on-black contours traced at 100, 130 and 
160~mK delineate the total power appearance of G152.4$-$2.1 at 21~cm. Overlaid 
on each map are thick pure white contours that outline HISA clouds seen in CGPS 
\ion{H}{i} maps at the same velocities. A single T$_{\textrm{B}}=$90~K contour 
(very thin black) outlines the brightest \ion{H}{i} cloud in the field, 
appearing in between the molecular clouds that follow the outside of the 
southern shell. All of this activity peaks at $v_{LSR}=-$12~km~s$^{-1}$.}
\label{g152_co}
\end{center}
\end{figure*}

Figure~\ref{g152_co} is a 9-channel montage of $^{12}$CO and \ion{H}{i} towards 
G152.4$-$2.1. Each channel is spaced to 0.82~km~s$^{-1}$ and the velocity 
resolution is 1.32~km~s$^{-1}$. The CO maps have been smoothed to 
1$\arcmin$ spatial resolution to match the \ion{H}{i}. In all velocity 
channels, line emission towards G152.4$-$2.1 shows no obvious cavity nor shell 
of neutral or molecular gas surrounding it \citep[as an SNR evolving in a 
stellar wind bubble might; e.g. 3C434.1,][]{fost04}. Rather, the continuum 
boundary is sharply delimited by a wall of bright CO that systematically wraps 
the Eastern boundary of the SNR from the South to the North. This bright CO 
(T$_{\textrm{A}}=$1-2~K) emission begins outside of the Southern shell boundary 
at $v_{LSR}=-$8.9~km~s$^{-1}$ and in each of the following channels ($-$10.5 to 
$-$13~km~s$^{-1}$) the CO neatly matches the curvature of the sharp South-East 
corner of the shell. Also tracing the SNR's South-East and East periphery are 
numerous \ion{H}{i} Self-Absorption (HISA) clouds (white contours in 
Fig.~\ref{g152_co}) in CGPS maps at the same velocities. The CO and HISA 
environment seems to form a {}``wall'' of cool dense gas on the sky next to the 
East continuum boundary of the SNR. As well, CO and HISA appear to neatly fill 
in an Eastern {}``bay'' formed by the continuum contours at $\ell=$152$\fdg$9, 
$b=-$1$\fdg$9. An isolated bright cloud of \ion{H}{i} emission (single 90~K 
thin black contour in each map of figure), the brightest such cloud in the 
entire field and the only \ion{H}{i} emission feature possibly associated with 
the SNR, is seen exactly in the gap between the CO clouds that wrap along the 
outside of the Southern shell, at $\ell=$152$\fdg$2, $b=-$2$\fdg$6. All of 
these spatial coincidences occur in nine channels between $-$8.9 and 
$-$15.5~km~s$^{-1}$. Their spatial relation with respect to the shell is 
similar in channels $-$10.5 to $-$13~km~s$^{-1}$, suggesting a systemic 
velocity of $v_{LSR}\simeq-$12$\pm$2~km~s$^{-1}$ for G152.4$-$2.1.

\begin{figure*}[!ht]
\begin{center}
\includegraphics[scale=0.85]{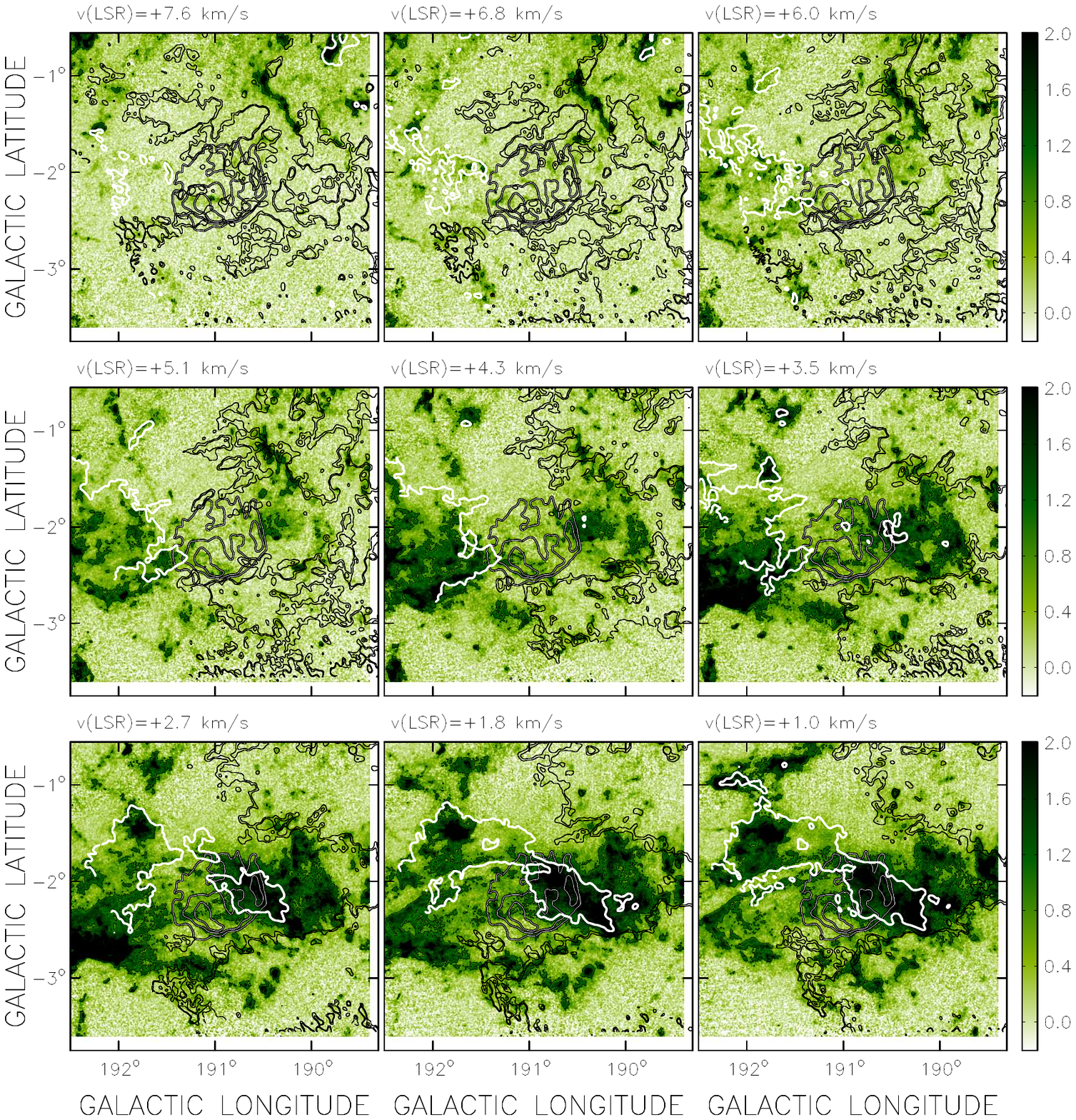}\vspace{-6cm}
\caption{$^{12}$CO line channel maps towards G190.9$-$2.2. The brightness scale 
is antenna temperature T$_{\textrm{A}}$ in K, and the resolution 1$\arcmin$. 
Three thick white-on-black contours traced at 100, 150 and 200~mK delineate the 
total power appearance of G152.4$-$2.1 at 21~cm. Extended HISA is contoured with 
thick pure white lines (level $-$20~K below the mean T$_{\textrm{B}}$ level of 
each \ion{H}{i} channel), and \ion{H}{i} emission at 6 and 9 Kelvin (above the 
mean of each map) is shown with the thin black contours. In channels with 
$v_{LSR}=+$4.3 and 5.1~km~s$^{-1}$ molecular gas is seen bracketing the SNR's 
East and West boundaries, and curved shell-like atomic gas clouds bracket the 
North and South boundaries.
}
\label{g190_co}
\end{center}
\end{figure*}

Figure~\ref{g190_co} is a 9-channel montage of $^{12}$CO and \ion{H}{i} centred 
on G190.9$-$2.2. Two bright CO clouds that begin to appear at 
$+$6.0~km~s$^{-1}$ flank the SNR on its left (East) and right (West) sides. 
This relationship is particularly striking in channels from $+$5.1 to 
$+$3.5~km~s$^{-1}$ where the CO clouds are dense and bright and continue to 
flank the SNR's East and West continuum boundary. At the same velocities an 
extensive HISA cloud that well traces the outline of the CO cloud in the East 
appears, and another HISA cloud just begins to appear atop the Western CO at 
$+$3.5~km~s$^{-1}$. In later channels $v_{LSR}\leq$2.7~km~s$^{-1}$ this 
CO+HISA appears atop the SNR's face, suggesting it is in front or behind it and 
sporting little relation with the continuum appearance. \ion{H}{i} emission 
surrounding G190.9$-$2.2 is much more complex than seen towards G152.4$-$2.1, 
and appears in the same velocity channels as the CO but in the opposite 
corners: above and below the object's North and South boundaries respectively. 
Both of these North and South \ion{H}{i} {}``caps'' follow the curvature of 
the SNR's limb at 21~cm in four channels $+$7.6 - $+$5.1~km~s$^{-1}$; 
thereafter the North arc disappears and the South has grown into a large 
\ion{H}{i} wall off of the South South-West corner of the SNR. The overall 
appearance of G190.9$-$2.2 being nestled within CO+HISA walls to the East and 
West and \ion{H}{i} emission shells to its North and South is particularly 
striking in the channel map at $v_{LSR}=+$5.1$\pm$1.6~km~s$^{-1}$. Based on 
this appearance we assign this as the systemic velocity of G190.9$-$2.2.

Even with a systemic velocity in hand, ascribing distances to these two objects 
is not straightforward. Under undisturbed circular rotation 
$v_{LSR}=-$12$\pm$2~km~s$^{-1}$ places G152.4$-$2.1 within the {}``Local arm'' 
at a distance of $\sim$1.1$\pm$0.1~kpc \citep[using the rotation curve 
of][]{fost10}. This SNR then has a linear diameter of 32$\times$30~pc. The 
gradient $\Delta r/\Delta v_{LSR}$ here is 110~pc per km~s$^{-1}$, which gives 
the \textit{minimum} uncertainty. However the association with HISA implies 
that neutral hydrogen at two different distances is present in these velocity 
channels (i.e. a cold foreground and a warm background), showing that circular 
rotation in this direction is anything but undisturbed and that the error on 
the distance is expected to be greater. 

A kinematic distance to G190.9$-$2.2 of 1.0$\pm$0.3~kpc is indicated based on 
the CO+\ion{H}{i}-based velocity of $+$5.1~km~s$^{-1}$, implying physical 
dimensions of 18$\times$16~pc. The uncertainty in this distance is large 
however, as the 1/sin$\ell$ projection here causes a small velocity range to 
represent a large line-of-sight path, and small velocity deviations from 
circular rotation are magnified into large distance uncertainties (here 
$\Delta r/\Delta v_{LSR}\sim$200~pc per km~s$^{-1}$). We know such deviations 
are present as remarkable and extended HISA is seen through the field, so
again the distance error is a lower limit.

\section{Conclusion and Future}
The purpose of this paper is twofold. Firstly we introduce two extended faint 
discrete objects discovered in the CGPS and show evidence (mainly through their 
radio spectral and polarization properties) that classifies them as SNRs.
Secondly we provide a basic qualitative interpretation and quantitative 
physical properties (fluxes, distances, sizes) that will aid the community in 
planning future new observations of them at radio and especially other 
wavelengths (e.g. X-ray, optical). The objects were discovered by deep point 
source subtraction (down to 2$\sigma$ level) and subsequent smoothing of two 
selected CGPS 21~cm continuum mosaics, to increase the S/N of faint extended 
emission. Their identity as SNRs was then discovered spectrally through 
comparison with other wavelengths (e.g. 74~cm component of the CGPS; the 
Sino-German 6~cm polarization survey). Undoubtedly many more such extended low 
surface brightness SNRs await discovery within this rich dataset using this 
approach. Such discoveries will further address the {}``missing SNR'' problem 
that is a key driver of new and planned deep radio surveys of the Milky Way's 
ISM.

\begin{acknowledgements}
We thank the referee for their thorough reading of our manuscript and the
many thoughtful suggestions they provided. We would also like to thank Chris 
Brunt (Exeter) for providing the CO data, and Dr. JinLin Han and Dr. Xuyang Gao 
(NAOC) for providing the 6~cm data prior to its public release. The Dominion 
Radio Astrophysical Observatory is a National Facility operated by the National 
Research Council. The Canadian Galactic Plane Survey is a Canadian project with 
international partners, and is supported by the Natural Sciences and 
Engineering Research Council (NSERC). The discovery of G190.9$-$2.2 by BC was 
made possible by a Brandon University Research Committee (BURC) grant to TF. 
The discoveries described here were made in part based on observations with the 
100-m telescope of the MPIfR (Max-Planck-Institut f\"ur Radioastronomie) at 
Effelsberg.
\end{acknowledgements}

\end{document}